\documentclass[a4paper,11pt]{article}
\usepackage{jheppub} 

\usepackage{xcolor}

\title{\boldmath Strategic White Paper on AI Infrastructure for Particle, Nuclear, and Astroparticle Physics: Insights from JENA and EuCAIF }

\affiliation[*]{Convener} 

\author[a,b]{Sascha Caron,\textsuperscript{*}} \affiliation[a]{IMAPP, Radboud University, Nijmegen, The Netherlands}
\affiliation[b]{Nikhef,
Science Park, Amsterdam, The Netherlands}

\author[c]{Andreas Ipp,\textsuperscript{*}} \affiliation[c]{Institute for Theoretical Physics, TU Wien, Wiedner Hauptstraße 8-10, 1040 Vienna, Austria}

\author[d]{Gert Aarts,} \affiliation[d]{Department of Physics, Swansea University, SA2 8PP, Swansea, United Kingdom}

\author[e,f]{Gábor Bíró,}
\affiliation[e]{HUN-REN Wigner Research Centre for Physics, 29--33 Konkoly--Thege Mikl\'os \'ut, H-1121 Budapest, Hungary}
\affiliation[f]{E\"otv\"os Lor\'and University, Institute of Physics and Astronomy, 1/A P\'azm\'any P\'eter s\'et\'any, H-1117 Budapest, Hungary}

\author[g,h]{Daniele Bonacorsi,}
\affiliation[g]{Physics and Astronomy Department (DIFA), Alma Mater Studiorum- Università di Bologna ,Italy} 
\affiliation[h]{INFN Sezione di Bologna Viale C. Berti Pichat 6/2 - 40126 Bologna, Italy}

\author[g,h]{Elena Cuoco,}

\author[i]{Caterina Doglioni,}
\affiliation [i] {University of Manchester, Oxford Road, M13 9PL, Manchester, United Kingdom}

\author[j,k]{Tommaso Dorigo,}
\affiliation [j] {Lule\aa \, University of Technology, Laboratoriev\"agen 14, Lule\aa, Sweden}
\affiliation[k] {Istituto Nazionale di Fisica Nucleare, sezione di Padova, via F. Marzolo 8, 35131 Padova, Italy}

\author[l]{Julián García Pardiñas,}
\affiliation[l]{Laboratory for Nuclear Science, Massachusetts Institute of Technology (MIT), 77 Massachusetts Ave, Cambridge, MA 02139, USA}

\author[m]{Stefano Giagu,}
\affiliation[m]{Department of Physics, Sapienza Universit\`a di Roma and INFN Roma, P.le A. Moro 5, 00185, Roma, Italy}

\author[n]{Tobias Golling,}
\affiliation[n]{Département de Physique Nucléaire et Corpusculaire, Université de Genève, Genève, Switzerland}

\author[o]{Lukas Heinrich,}
\affiliation[o]{Physics Department, Technical University of Munich, 80233 Munich, Germany}

\author[p]{Ik Siong Heng,}
\affiliation[p]{School of Physics and Astronomy, University of Glasgow, G12 8QQ, United Kingdom}

\author[q]{Paula Gina Isar,}
\affiliation[q]{Institute of Space Science - INFLPR Subsidiary, Atomistilor 409, 077125 Magurele, Ilfov county, Romania}

\author[r]{Karolos Potamianos,}
\affiliation[r]{Physics Department, University of Warwick, Coventry CV4 7AL, United Kingdom}

\author[s]{Liliana Teodorescu,}
\affiliation[s]{Brunel University of London, Kingston Lane, UB8 3PH, Uxbridge, United Kingdon}

\author[p]{John Veitch,}

\author[t]{Pietro Vischia,}
\affiliation[t]{Universidad de Oviedo and ICTEA,  Oviedo, Principado de Asturias, España}

\author[u]{Christoph Weniger}
\affiliation[u]{GRAPPA, Institute of Physics, University of Amsterdam, Science Park, Amsterdam, The Netherlands}

\emailAdd{scaron@nikhef.nl}
\emailAdd{andreas.ipp@tuwien.ac.at}

\abstract{Artificial intelligence (AI) is transforming scientific research, with deep learning methods playing a central role in data analysis, simulations, and signal detection across particle, nuclear, and astroparticle physics. Within the JENA communities—ECFA, NuPECC, and APPEC—and as part of the EuCAIF initiative, AI integration is advancing steadily. However, broader adoption remains constrained by challenges such as limited computational resources, a lack of expertise, and difficulties in transitioning from research and development (R\&D) to production. This white paper provides a strategic roadmap, informed by a community survey, to address these barriers. It outlines critical infrastructure requirements, prioritizes training initiatives, and proposes funding strategies to scale AI capabilities across fundamental physics over the next five years.
\\
\vspace{2em}
\\
Version:  2025-03-12
}

\begin{document}
\maketitle
\flushbottom

\newpage

\section*{Executive Summary}
\addcontentsline{toc}{section}{Executive Summary} 

Advances in artificial intelligence (AI) are transforming fundamental physics research across the JENA communities (ECFA, NuPECC, APPEC). This white paper presents 12 strategic recommendations to scale AI capabilities, addressing challenges such as resource limitations, integration, and training gaps.
These investments will also strengthen expertise in this important technology in Europe, ensuring long-term benefits beyond fundamental physics.

\begin{itemize}
\item 
(R1) Convene dedicated discussions with national research groups and funding bodies to assess and compare the feasibility of a \textbf{centralized large-scale GPU facility versus   federated and hybrid high-performance computing (HPC) infrastructures}, supported by working groups developing detailed implementation plans for both options, with the aim of accelerating the deployment of a scalable AI infrastructure.

\item
(R2) Establish a \textbf{scalable data infrastructure} initiative by creating shared repositories and tools, and \textbf{developing platforms for distributed workloads}. These efforts need targeted funding programs and a concrete community-driven structure to ensure widespread adoption  and collaboration in AI research.

\item
(R3) Encourage funding to \textbf{transition AI-driven R\&D activities into pro\-duct\-ion-ready applications} within established experimental workflows, focusing on adopting best practices to achieve practical, scalable improvements without requiring a complete system overhaul.

\item
(R4) Allocate \textbf{dedicated funding to establish and support specialized Machine Learning Operations (MLOps) personnel} to streamline the integration and ensure the sustainable maintenance of AI models within production workflows. This effort should encompass the development of community-wide standards, tools, and platforms to effectively manage the entire lifecycle of machine learning models.

\item
(R5) Invest in the \textbf{creation of ``science Large Language models (LLMs)''} tailored to the unique challenges of fundamental physics and science, balancing the use of commercial tools for general tasks with specialized models for domain-specific needs. This requires dedicated funding, access to large-scale GPU infrastructure, and collaborative frameworks to enable transparent, efficient, and impactful AI solutions.

\item 
(R6) Establish dedicated funding schemes and a collaborative structure to develop community-driven \textbf{foundation models trained on domain-specific data to learn meaningful representations serving a large variety of downstream tasks}.
This effort should identify representative benchmarks, extendible in complexity and realism by  integrating both synthetic and real-world data to address domain-shift issues, leverage physics-informed augmentations, ensure models are rooted in scientifically relevant tasks, and foster automation, explainability and interpretability to accelerate AI advancements in the field, and to develop a well-defined AI demonstrator for the wider AI community.

\item 
(R7) Establish a dedicated effort to \textbf{develop and maintain extensible benchmarks for various AI tasks in fundamental physics}, such as event classification, parameter inference, tracking and anomaly detection. Support efforts to encourage researchers to share well-documented surrogate models to promote reusability and collaboration to drive innovation and standardisation in this area.

\item 
(R8) Investigate and adopt benchmarks that are suitable for fundamental sciences to raise \textbf{awareness of the environmental impact} of  large AI models. 
Consider collective mitigation strategies such as optimising widely used frameworks and models and their interfaces to existing software frameworks, as well as individual strategies that lead to minimal/acceptable performance loss. 
Cooperate with infrastructure and computing sites to minimise  carbon costs of compute-intensive AI tasks.

\item 
(R9) Develop activities aiming to \textbf{integrate FAIR compliance into publication criteria and practices}, recognise and incentivise the FAIR compliant work in policy and funding measures as well as career progression, build community awareness through training and collaboration, and support the development of technical tools and standards to facilitate the adoption of the FAIR principles.

\item 
(R10) Fund the \textbf{development and organization of practical 
training courses and summer schools} to equip researchers with the skills to implement open research and reproducibility requirements, incorporating examples and industry perspectives. Facilitate partnerships with industry to sponsor training events and provide placements for early-career and senior researchers, enhancing their AI and data science expertise while fostering connections between fundamental science and commercial applications.

\item 
(R11) \textbf{Establish interdisciplinary research initiatives} that bring together physicists, AI specialists, software engineers, HPC experts, and potentially experts from other related fields, to tackle large-scale projects. Provide dedicated funding to support \textbf{cross-domain knowledge transfer} through workshops, training programmes and open source collaboration. Invest in shared repositories and computing platforms to enable data sharing, modelling development and collaboration between different disciplines.

\item 
(R12) \textbf{Establish and support a dedicated organisational structure} to coordinate strategic investments in AI for fundamental physics to accelerate the development and deployment of innovative AI technologies tailored to the specific challenges of the field. Existing initiatives like the European Coalition of AI for Fundamental Physics (EuCAIF) can serve as a model for such efforts.
\end{itemize}

\newpage 

\section{Introduction}
\label{sec:intro}

Over the last decade, artificial intelligence (AI) has witnessed significant advancements, particularly in the domains of machine learning (ML) and its subset, deep learning (DL), which have transformed numerous applications.
These developments are already
transforming scientific research and discovery.  State-of-the-art algorithms such as deep artificial neural networks, generative models and large-scale language models have already opened up new possibilities, including application to scientific research.
Within the JENA communities---ECFA (European Committee for Future Accelerators), NuPECC (Nuclear Physics European Collaboration Committee), and APPEC (Astroparticle Physics European Consortium)---AI has begun to transform research methodologies, from particle physics to astroparticle physics and nuclear physics.
In addition, EuCAIF (European Coalition for AI in Fundamental Physics) \cite{eucaif} has emerged as a new initiative to drive forward AI innovations in these areas. The application of AI technologies in these fields has shown enormous potential, and DL methods are now used routinely in signal detection, pattern recognition, parameter estimation, model selection and simulations.  
As fundamental physics advances into a new era of big data and complex simulations, the integration of AI is emerging as a critical driver of discovery and innovation. Review articles on AI applications in particle physics \cite{Albertsson:2018maf,Radovic2018}, nuclear physics \cite{Boehnlein:2021eym}, and gravitational wave science \cite{Cuoco:2020ogp} offer  insights into the  potential of these technologies.

Although modern DL methods are already used in production, \textit{i.e.}, in the standard software pipelines of large experiments such as ATLAS and CMS and---so far less so---in the analysis chains of smaller experiments in nuclear and astroparticle physics, the use of state-of-the-art  AI methods is often limited to development and small research projects. 
In the last ten years, there has been a great deal of research and development work, in which the application of new ML methods in fundamental physics have been proposed and investigated, and a variety of ML models, from supervised learning algorithms to sophisticated generative models, are being tested on experimental data. 
However, these efforts remain relatively scattered, and  are often hampered by the availability of graphics processing units (GPUs), limited high-performance computing (HPC) resources, the slowness of integrating tools from R\&D to production, maintaining a large number of trained ML models in production, and the requirements of traditional grid and cloud-based data processing infrastructures. Furthermore, benchmarks for ML performance in scientific contexts remain underdeveloped, with most efforts focused on improving software tools that meet the particular computational requirements of physics experiments. 

Another key challenge for the adoption of AI in fundamental physics lies in the need for specialized education and training. While there are numerous online courses and workshops, the expertise required to apply these technologies effectively within the context of physics remains rare. At present, physicists with ML skills are largely self-taught or trained in an ad-hoc fashion through collaborations. Organizational structures within the JENA communities are beginning to address these gaps, with institutions offering dedicated ML training programs for early-career researchers and established scientists alike. However, there is still a pressing need for a more formal and systematic approach to building AI competency within the field, from curriculum development to hands-on experience in real-world research projects. Furthermore, when it comes to developing large-scale AI models,  such as foundation models, there is no organizational structure or structural funding. 

This white paper is intended to provide a strategic overview of AI infrastructure needs for fundamental physics and provide guidance for future funding and development efforts. Based on input from the JENA communities (a survey), this document sets out the recommendations for scaling up AI capabilities over the next five years.
An overview as part of the US Community Study on the Future of Particle Physics (Snowmass) can be found in Refs.~\cite{Shanahan:2022ifi,Boyda:2022nmh}.

\section{Survey: Community Insights}

\subsection{Overview of the Survey}
To develop a comprehensive understanding of the community's views, we conducted a survey among researchers in high-energy physics, nuclear physics, astrophysics and the (related) HPC community. The questionnaire aimed to gather insights into the current infrastructure and future needs regarding AI  usage in fundamental physics research. The survey was distributed to the ECFA, NuPECC and APPEC mailing lists and mailing lists from EuCAIF. Additional calls to HPC groups expanded the participation. 
A total of 137 responses were collected from July to November 2024.
The survey included 40 questions, including ones on the use of AI tools, available computational infrastructure, and the anticipated future requirements for AI development.

\begin{figure}[h!]
    \centering
    \includegraphics[width=0.8\textwidth]{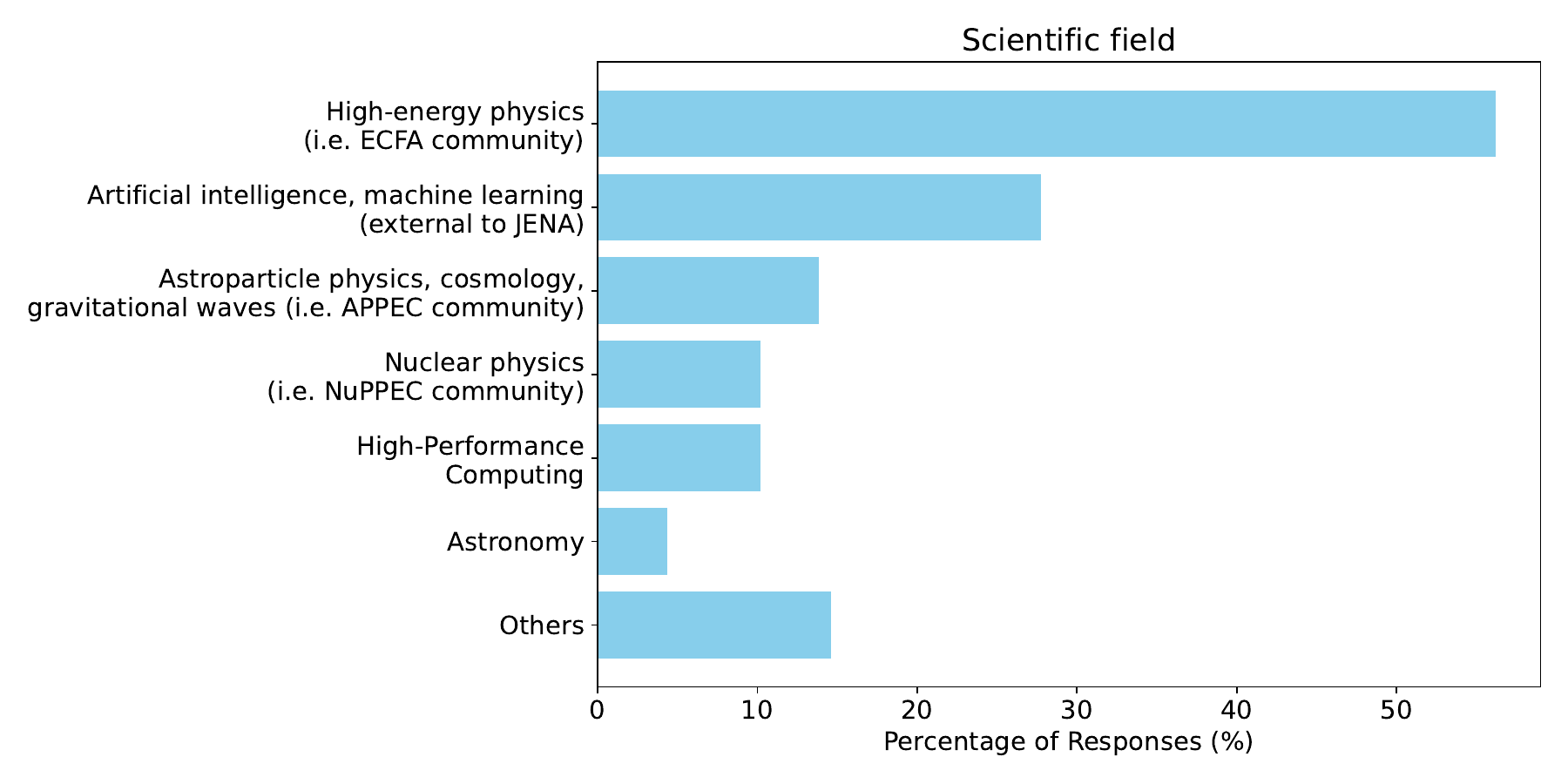}
    \caption{Response to the question ``[2/40] What is your main scientific field (or JENA community)?''}
    \label{fig:q2}
\end{figure}

Most respondents were individual researchers (70\%) or group leaders (26\%). 
A large fraction of the respondents (84\%) have a European affiliation. 
A significant portion of respondents came from the high-energy physics (ECFA) community (56\%) and AI communities external to JENA (28\%), astroparticle physics (APPEC, 14\%), nuclear physics (NuPPEC, 10\%), and HPC centers (10\%) (see Fig.~\ref{fig:q2}). 
 It is worth noting that many
participants, particularly PhD students, may not have been familiar with the affiliation with JENA,
which may influence the representation in the category ``AI outside JENA.''

\subsection{Key Findings}

\begin{figure}[h!]
    \centering
    \includegraphics[width=0.8\textwidth]{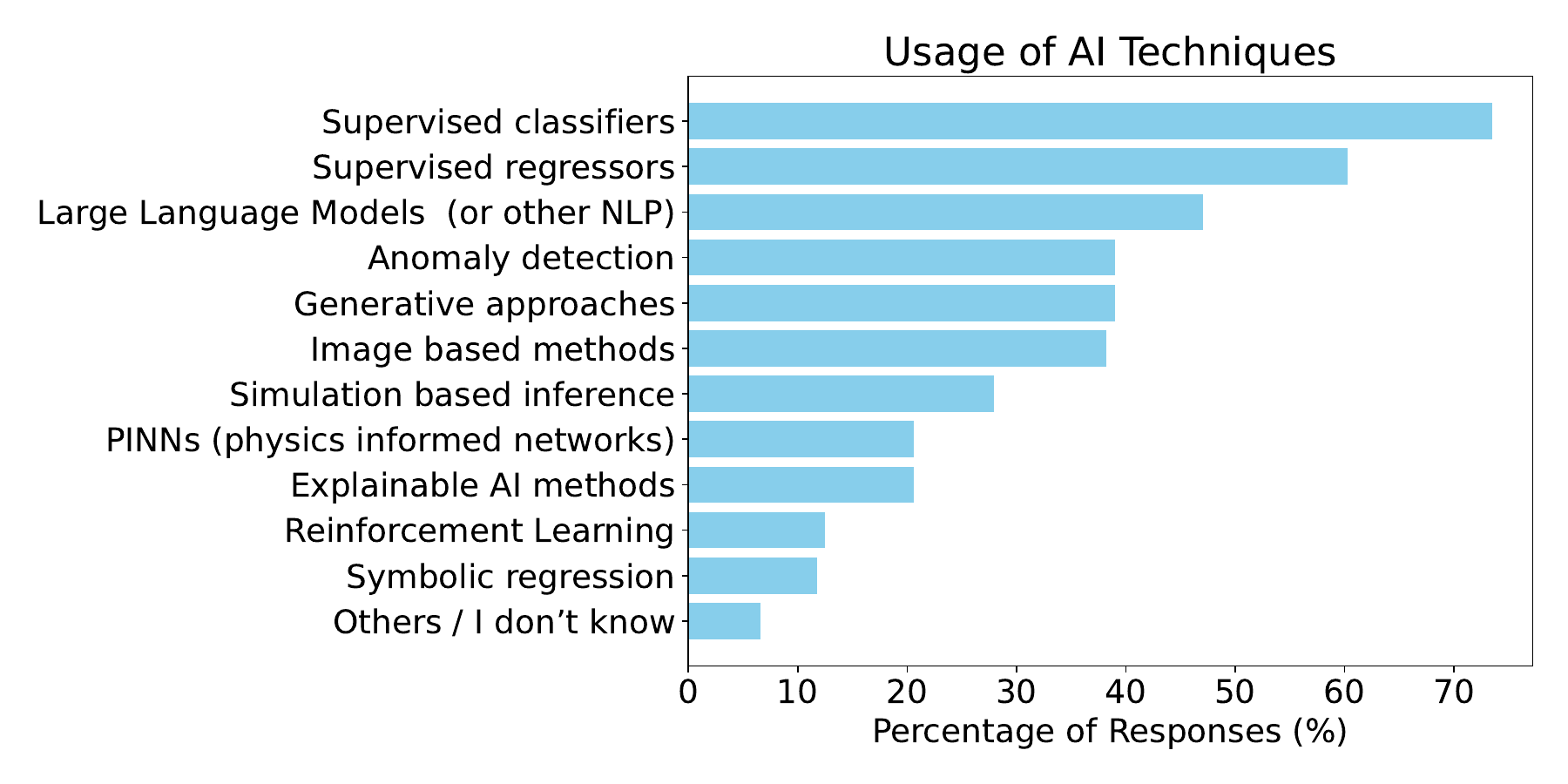}
    \caption{Response to the question ``[5/40] What is your usage of AI? Which ML techniques do you use?'' }
    \label{fig:q5}
\end{figure}

Tools such as ChatGPT and other large language models (LLMs) have become ubiquitous, with 74\% of respondents using commercial tools, and 26\% having also worked with open-source alternatives. 
However, 20\% of the community does not use these tools at all. Beyond 
conversational AI, machine learning in scientific research covers a much wider range of applications. Researchers employ ML for tasks such as data analysis, model development, and predictions across diverse fields of high-energy and nuclear physics. Commonly used techniques include supervised learning, generative approaches, and to lesser extent symbolic regression (see Fig.~\ref{fig:q5}). 
The data types used range from sensor or detector data, 4-vectors, time series, images, or text for offline processing, while only a small fraction (16\%) work with real-time data. 
The most commonly used frameworks for training algorithms are the DL libraries PyTorch (71\%) and TensorFlow (44\%), as well as the traditional ML library Scikit-learn (49\%).

Researchers in the field rely on a wide range of computational resources to meet the diverse requirements of their ML workloads. They make equal use of local installations (laptops or desktop computers), local computing farms, and large-scale computing clusters (55\%) for running ML workloads. 
While most installations are local, containers as virtualization solutions are used to a lesser extent. The size of training data for ML tasks varies widely, with over half the use cases in the 10 GB to 1 TB range. The total hours of compute resources on CPU and GPU, including training, testing, and inference, span a broad spectrum from less than 10 hours to more than a million hours, evenly distributed on a logarithmic scale. The majority of machine learning workloads are executed on CPUs, consumer GPUs, and enterprise GPUs, while only a small fraction experiment with FPGAs, TPUs, or other specialized hardware.  Most applications use a single GPU card or a node with up to 8 GPUs, with only about 15\% of users requiring multiple nodes, each equipped with several GPUs.
GPU memory requirements also vary significantly, from 16 GB to 100 GB, with only 5\% of users currently needing more than 100 GB of GPU memory. 
A similar distribution can be observed for host system memory, with around 12\% of users requiring more than 128 GB of CPU memory. 
In summary, the community employs a variety of hardware configurations, from consumer-grade setups to large-scale clusters, depending on the specific demands of their ML tasks.

\begin{figure}[h!]
    \centering
    \hfill \includegraphics[width=0.4\textwidth]{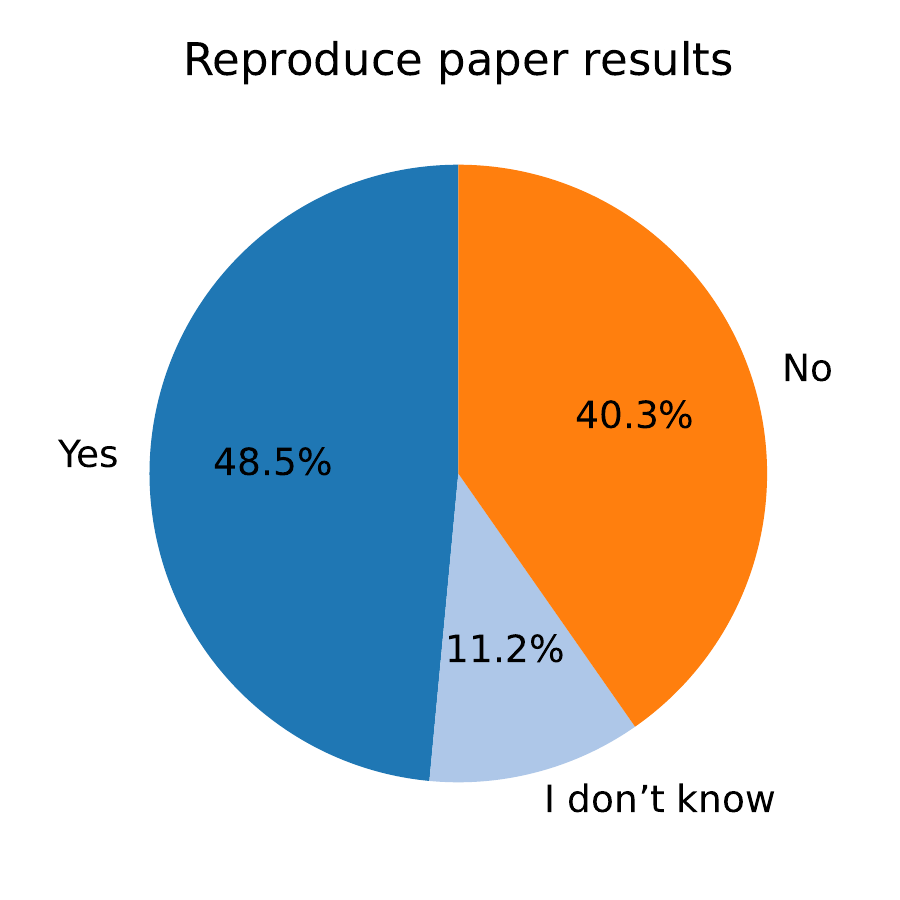}
    \hfill\includegraphics[width=0.4\textwidth]{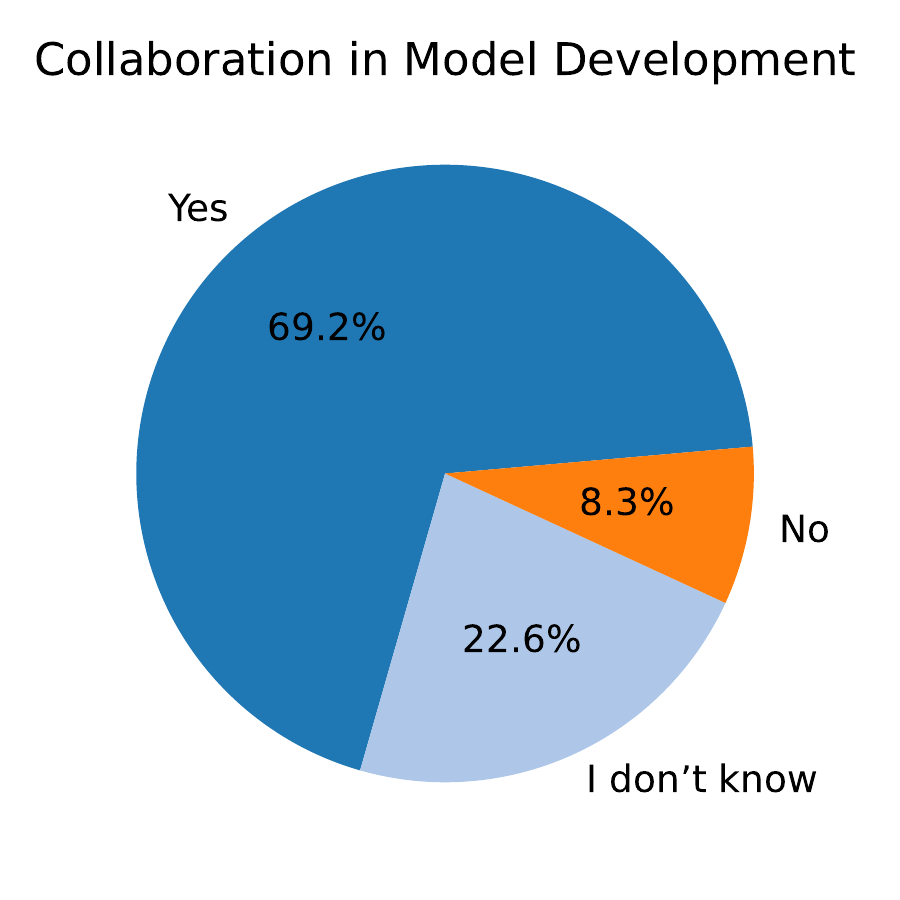} \hfill \,
    \caption{Response to the questions ``[20/40] Did you ever manage to reproduce someone’s paper results / retrain their model?'' (left panel) and ``[30/40] Should we collaborate more in the development of large-scale ML models (e.g.~foundation models) for physics?'' (right panel).}
    \label{fig:q20}
    \label{fig:q30}
\end{figure}

Several challenges have been identified within the community, with the most pressing being the lack of access to computational resources and the availability of specialized hardware. Almost half of the respondents reported successfully reproducing someone’s paper results (see Fig.~\ref{fig:q20}). 
For those who could not, the primary reasons were inadequate documentation and the lack of well-functioning software workflows. Additionally, the lack of human resources was frequently cited as a barrier to reproducibility and scalability. Interestingly, only a small fraction (16\%) of respondents indicated that limited computational resources were the main issue.

\begin{figure}[h!]
    \centering
    \hfill \includegraphics[width=0.48\textwidth]{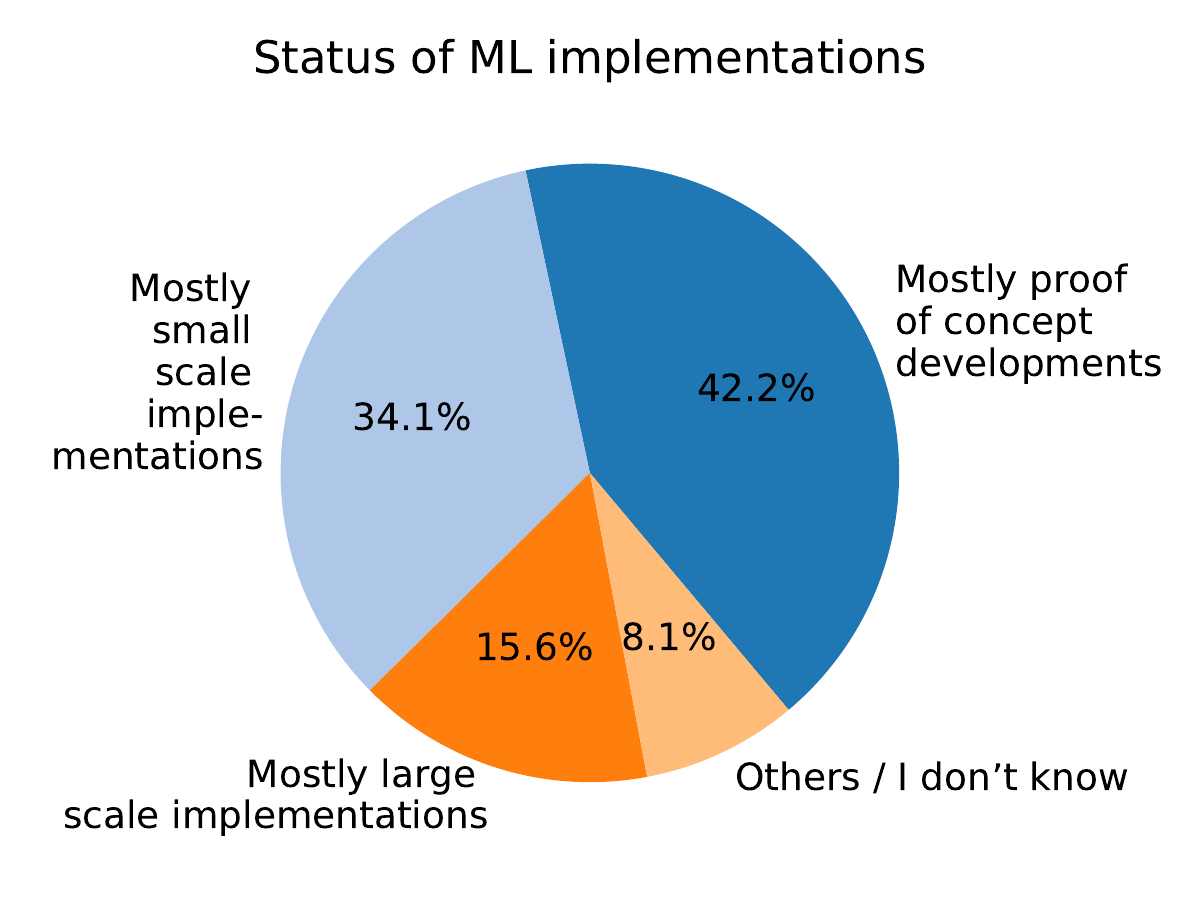}
    \hfill\includegraphics[width=0.48\textwidth]{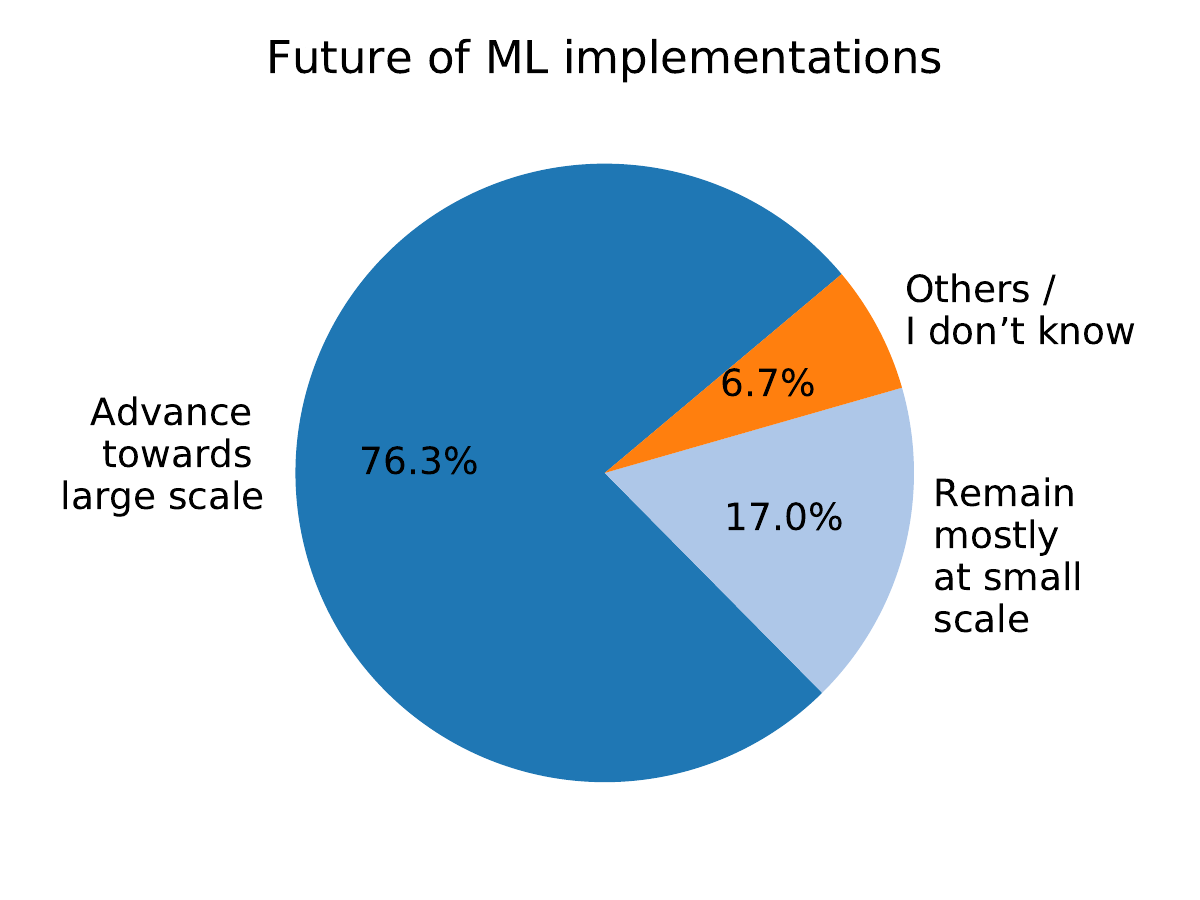} \hfill \,
    \caption{Response to the questions ``[9/40] In your current work, what is the predominant status of your machine learning implementations?'' (left panel) and ``[34/40] How do you foresee the status of ML implementations evolving in your field over the next five years?'' (right panel).}
    \label{fig:q9}
    \label{fig:q34}
\end{figure}

Another challenge lies in the current usage patterns: most machine learning implementations remain at the proof-of-concept or small-scale stage, with large-scale applications being the exception (see Fig.~\ref{fig:q9}). Scaling these smaller efforts to larger systems is seen as a significant hurdle for the community.

Looking ahead, the community’s primary needs revolve around training, 
enhanced access to high-end GPUs and improvements in scalability for ML applications. A significant number of respondents emphasized the importance of developing common representative and reproducible benchmarks, which would provide a standardized way to measure progress and compare methods. Additionally, the need for dedicated frameworks to support training ML models was highlighted as crucial by many researchers.
55\% of respondents support the idea of an organizational effort to facilitate this. 
EuCAIF could play a central role in collecting and designing public benchmarks and datasets.

There is growing recognition that large-scale ML models, including LLMs, will be pivotal in the future of research. 63\% of respondents consider large-scale ML models important for advancing physics, 
and 69\% agree on the need for collaboration in developing these models (see Fig.~\ref{fig:q30}). 
Specifically, 58\% support the creation of a domain-specific LLM for fundamental physics, potentially trained on datasets from repositories like arXiv’s High Energy Physics (hep) and Astrophysics (astro). 
Many researchers believe that trends like transformers and LLMs are already driving the need for large-scale GPUs in the community. 

\begin{figure}[h!]
    \centering
    \includegraphics[width=0.75\textwidth]{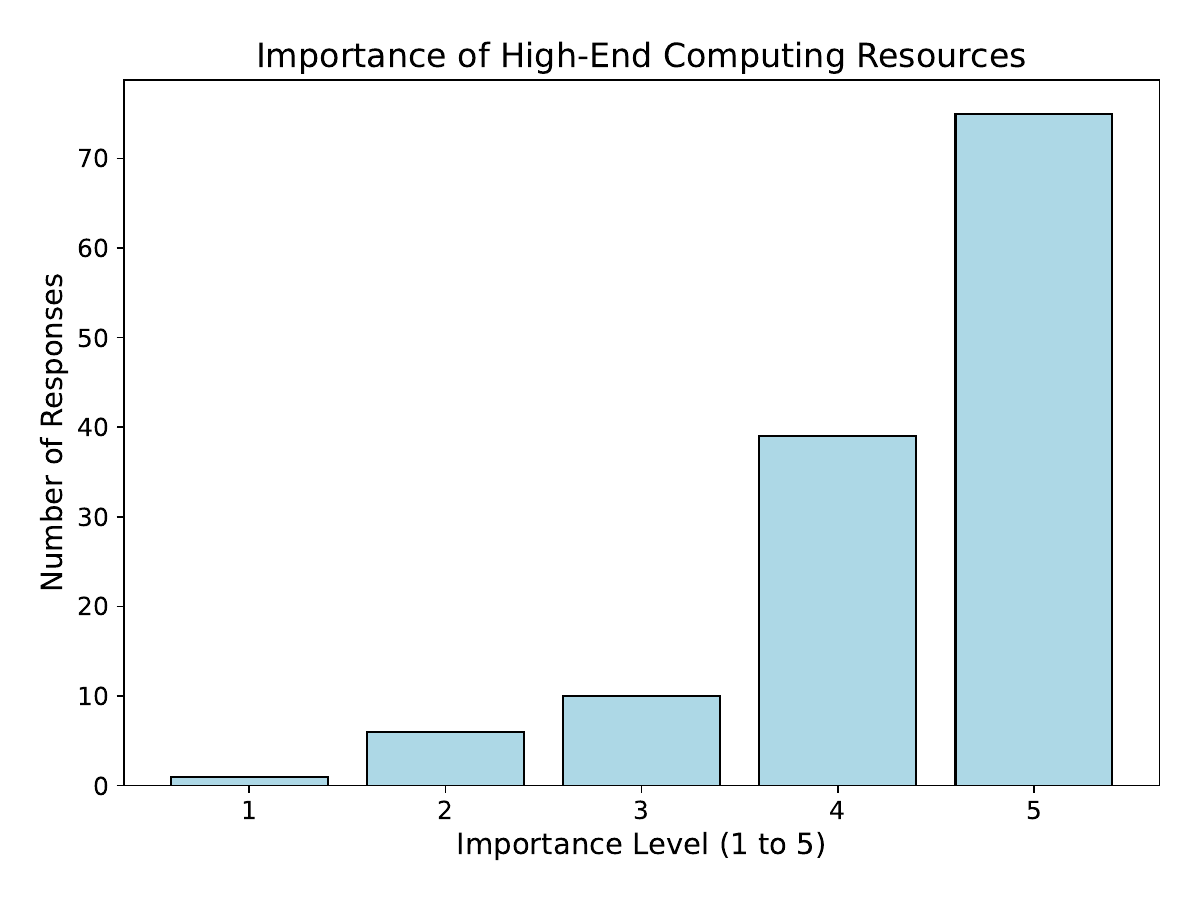}
    \caption{Response to the question ``[26/40] How important would high-end/large GPUs (like the A100 etc) be for training large-scale deep learning models over the next five years?'''}
    \label{fig:q26}
\end{figure}

As research continues to grow in complexity, interactive platforms such as Google Colab and JupyterHub are expected to play an increasingly important role in ML workflows. The community anticipates that compute resource requirements for ML tasks will continue to rise, with high-end GPUs (like NVIDIA A100) becoming essential for training deep learning models (see Fig.~\ref{fig:q26}).
Moreover, three quaters of respondents predict that ML implementations will evolve from mostly small-scale or proof-of-concept stages to large-scale applications over the next five years (see Fig.~\ref{fig:q34}). To support these growing needs, 76\% of respondents believe that a large-scale distributed GPU facility would significantly benefit their research, underlining the necessity for more centralized and accessible computing resources.

\subsection{Interpretation of Results}

The results of the survey reveal broad engagement with AI across the community, but they also highlight a stark divide between the current usage patterns and the needs for future growth. While there is widespread adoption of DL tools like PyTorch and TensorFlow, the survey indicates that most applications are still in the proof-of-concept or small-scale phase. This suggests that the community is in an exploratory stage, where many researchers are experimenting with ML techniques but have not yet fully scaled their applications to meet the growing demands of their fields.

A clear consensus has emerged around the need for greater collaboration and infrastructure to support the transition to large-scale ML implementations. The call for standardized benchmarks, dedicated frameworks, and shared computational resources points to a community that is ready to move beyond isolated efforts and embrace more coordinated approaches. The strong support for creating domain-specific large-scale ML models and LLMs for physics further underscores this trend. Additionally, the data reveal an important shift in computing requirements, with the community anticipating significantly higher demand for GPU resources and distributed systems in the coming years. 

In summary, the survey points to a community that is eager to embrace the potential of AI but requires better access to resources, stronger collaborations, and scalable frameworks to fully realize this potential. These findings set the stage for the recommendations that follow, which will address the steps needed to bridge the gap between current capabilities and future aspirations.

\section{Key Messages and Recommendations}

\subsection{(R1) Scalability and Access to HPC Resources}  
The JENA community’s ability to conduct impactful AI research is tightly linked to the availability of scalable and accessible HPC resources. The survey conducted for this white paper reveals a strong demand for large, distributed GPU facilities, with nearly 75\% of respondents supporting such infrastructure. This underscores the pressing need to expand HPC resources to meet the growing computational demands of ML in large-scale modeling, foundational model training, and experiments with massive datasets, such as those anticipated from the High Luminosity Large Hadron Collider (which should be operational in mid-2030) and other experimental programs.
As AI becomes integral to research workflows, computational requirements for deep learning model training, parameter optimization, and large-scale simulations have surged. Traditional CPU-based infrastructures used in high-energy physics are insufficient to handle the massive data volumes generated in contemporary research. This requires a transition to GPU-based facilities that excel at intensive parallel computing tasks and AI models development.
To address these demands, the community strongly supports the establishment of large-scale, distributed GPU facilities. These would not only enhance computational capacity but also provide equitable access through centralized and distributed clusters. Key design considerations include high-bandwidth connectivity, low-latency access, and advanced scheduling algorithms to maximize resource utilization and enable efficient workload distribution.
The EuroHPC AI Factories funding initiative, as outlined in the European Commission's digital strategy \cite{eurohpc_ai_factories}, emphasizes the development of cutting-edge AI infrastructure and services and may provide an opportunity to address scalability challenges.
Two primary options emerge for addressing these scalability challenges:
\begin{itemize}
\item \textbf{Option 1:} establish a centralized, large-scale GPU facility that consolidates resources across countries and institutions. This would maximize GPU utilization for sporadic, resource-intensive workloads, like training foundation models, promote collaboration by easing data sharing and model development, and enable investment in experimental hardware impractical for individual institutions. Challenges include competition for resources, high costs and logistical hurdles of transferring large datasets, and the political and operational complexities of managing a shared international facility.
\item\textbf{Option 2:} expand existing HPC infrastructures across multiple institutions, by increasing GPU availability and integrating cloud-based solutions with on-premises systems. This hybrid model offers flexibility for scaling ML workloads dynamically, reduces data transfer delays, and allows institutions to retain control over their resources. However, this approach risks fragmentation, higher overall costs due to duplication of resources, and management complexity from disparate hardware and software configurations.
\end{itemize}
Both approaches align with the growing need for scalable and collaborative infrastructure. 

\textbf{Recommendation:}
Organize dedicated discussions to evaluate the two infrastructure options in detail. This should include: \textbf{1)} scheduling focused meetings with national research groups and funding councils to assess feasibility and resource availability; \textbf{2)} establishing working groups to develop detailed implementation plans for both options.
The community should aim to initiate these discussions as soon as possible, to evaluate the feasibility of each approach and to start the implementation of necessary AI infrastructure.

\subsection{(R2) Data Infrastructure and Infrastructure for Distribution}

The demand for improved data infrastructure and management is increasingly urgent as AI models continue to scale up: more than 76\% of the community foresee an advancement towards large scale implementations over the next five years in their respective fields. Community feedback also reveals that while 34\% of datasets used in ML model training range between 10-100 GB, 20\% exceed 100 GB, with some surpassing 1 TB, highlighting the necessity of robust data-sharing frameworks to accommodate these vast datasets. However, reproducibility remains a challenge: while 49\% of practitioners have successfully reproduced results, 40\% have not, largely due to inadequate documentation (55\%).

To address these gaps, adopting best practices from data science, such as structured data labeling and standardized documentation, could be transformative. 
A dedicated infrastructure for AI workload distribution is also essential to foster inclusivity, enabling smaller research groups with limited resources to participate more fully. Establishing common tools and frameworks could reduce the time wasted on duplicative efforts, streamline interoperability (for instance, via ONNXRuntime for ML inference), and accelerate scientific advancement. By building a supportive ecosystem with accessible resources and documentation, the community can advance the standards of data infrastructure, ensuring sustainable growth in AI and facilitating broader participation and reproducibility in research.

\textbf{Recommendation:}
Establish a scalable data infrastructure initiative by creating shared repositories and tools, and developing platforms for distributed workloads. These efforts need targeted funding programs and a concrete community-driven structure to ensure widespread adoption  and collaboration in AI research.

\subsection{(R3) Integrating AI from R\&D into Production-ready Applications}

Despite notable advancements, R\&D for AI in physics remains fragmented, with most efforts focused on proof-of-concept applications rather than integrated, production-ready solutions.
In particle physics, efforts towards deploying AI algorithms into the real-time processing for trigger and data acquisition are taking shape.
In astroparticle research, AI techniques, including the use of hardware accelerators and differentiable programming, are enabling the processing of large datasets from international collaborations, often with vast improvements in latency when pre-trained methods can be used.
Generative methods are accelerating costly simulations and inference, both as end-to-end replacements and via integration into existing pipelines.

While there are undoubtedly many efforts already embedded in the workflows of experimental collaborations, these efforts are carried out by a relatively small number of researchers, mostly due to challenges in devising (and implementing) robust solutions beyond proofs of concept (PoC).
Established large-scale analyses often carry a great deal of inertia and high requirements for reproducibility and interpretability impeding the adoption of new technology.
Much AI research in this space demonstrates promising concepts but often stops short of integrating these ideas into fully implemented experimental workflows.
Moreover, collaboration with AI practitioners outside of physics can be challenging due to the difficulty of sharing experimental data, which is often restricted by data embargoes and is highly specialised. 
Navigating physics data often involves a steep learning curve adjusting to historically grown work pipelines and limiting data formats.
Although open competitions may partially bridge the gap, these are simplified and certainly require a lot of effort to be implemented in experimental workflows.
Last but not least, there is a big discrepancy among researcher groups in terms of access to suitable compute to devise AI solutions, as only limited, but expanding, support is available to be shared within experimental collaborations (as opposed to institutional resources).

\textbf{Recommendation:} 
Encourage funding to transition AI-driven R\&D activities into production-ready applications within established experimental workflows, focusing on adopting best practices to achieve practical, scalable improvements without requiring a complete system overhaul.

\subsection{(R4) Advancing Machine Learning for Production Workflows}

A notable 15\% of respondents have already achieved large-scale ML implementations, highlighting the community's engagement with production-level challenges. Over the next five years, 77\% anticipate scaling their ML work to large production levels, 
which will introduce significant challenges in managing the amount and variety of models. Key concerns include versioning, bookkeeping, retraining as conditions change, software compatibility, and storing model weights and data for reproducibility. These challenges are addressed in MLOps (Machine Learning Operations), a set of practices that streamline the lifecycle of machine learning workflows, encompassing model development, deployment, monitoring, and maintenance. Standardisation and common tools are key, with 73\% of respondents highlighting the importance of shared ML frameworks. 
While industry standards should be adopted wherever possible, to profit from external knowledge and efforts to maximise efficiency, unique constraints in our field will likely require the development of custom community-level solutions at different levels. Survey suggestions include developing community-level platforms for model and data sharing, inference frameworks and (semi-)automated pipelines, together with associated documentation and guidelines. Where experiment-level specifications and regulations may limit the adoption of community-wide tools, said approaches can at least be applied at the experiment level.

The development and maintenance of such tools requires dedicated personnel. Respondents emphasised the need for specialised support roles, funding not tied to specific projects, and continued maintenance. Creating career paths for MLOps professionals is essential to sustain these efforts.

\textbf{Recommendation:}
Allocate dedicated funding to establish and support specialized MLOps personnel to streamline the integration and ensure the sustainable maintenance of AI models within production workflows. This effort should encompass the development of community-wide standards, tools, and platforms to effectively manage the entire lifecycle of machine learning models.

\subsection{(R5) Large Language Models and Fundamental Physics}
\label{sec:LLM}

With 70\% of survey respondents already using tools such as ChatGPT, it is important to weigh up both the benefits and limitations of these commercial AI models for fundamental physics. While such tools provide accessible, pre-trained models that increase the efficiency of scientists, they also pose a challenge. 
Their generality limits domain-specific accuracy and knowledge, especially in the complex contexts of fundamental physics (e.g.~knowledge of analysis at the LHC, FAIR or in gravitational wave experiments). In addition, dependence on external platforms raises serious concerns about privacy, control and long-term usability for research.
As we progress in AI research, the question arises: should we continue to rely on large commercial models or focus on developing and fine-tuning our own community-driven models tailored to scientific needs? 
Developing “science community models” would allow us to customize AI to our own data and problems, improving transparency and control. However, this path comes with significant technical requirements - in particular, access to large GPU infrastructure, dedicated funding and a collaborative structure needed to train and deploy these sophisticated models. The recommendations of the European strategy can play a serious and important role here.

To be successful, we need to invest in collaborative efforts to share data, resources and GPU capacity. Building such models within the community requires not only a state-of-the-art computing infrastructure, but also sustained collaboration between AI specialists and fundamental physics in general.
A balance between these options - using commercial tools for certain tasks and developing specialized models for others - will be crucial to fully exploit the potential of AI in fundamental physics.

\textbf{Recommendation:} 
Invest in the creation of ``science LLMs'' tailored to the unique challenges of fundamental physics and science, balancing the use of commercial tools for general tasks with specialized models for domain-specific needs. This requires dedicated funding, access to large-scale GPU infrastructure, and collaborative frameworks to enable transparent, efficient, and impactful AI solutions.

\subsection{(R6) Foundation Models and Large-scale Machine Learning}Foundation models (FMs) are large-scale ML models designed to serve as a backbone for a broad range of downstream tasks. These models are pre-trained on vast, diverse, and often multi-modal datasets. This allows them to learn representations or generate data that are broadly applicable across different domains. Once trained, FMs can be fine-tuned or adapted for specific tasks with relatively small amounts of additional data (post-training), making them versatile tools. FMs are characterized by: {(i) scale}, often comprising billions of parameters; {(ii) multi-modality}, they can process multiple data modalities individually or in combination; {(iii) pre- and post-training}, using large datasets in a self-supervised or weakly supervised manner for pre-training; and {(iv) broad applicability}, across a wide variety of tasks.

The LLM’s of the previous Section~\ref{sec:LLM} are FMs, trained domain-independently and fine-tuned to domain-specific data.  Here a more physics-specific definition is used: as flexible models trained for different tasks, but on domain-specific data, such as 4-vectors, detector hits or astrophysical images, with minimal fine-tuning.

FMs provide a unique opportunity to address the bottlenecks of the traditional analysis approach in HEP which currently requires vast amounts of synthetic data ({compute bottleneck}), it leaves little room for re-utilization due to highly specific approaches ({person-power bottleneck}), and the domain shifts between real and synthetic data lead to a {bottleneck of systematic uncertainties}.

\textbf{Recommendation:} 
Establish dedicated funding schemes and a collaborative structure to develop community-driven foundation models trained on domain-specific data to learn meaningful representations serving a large variety of downstream tasks.
This effort should identify representative benchmarks, extendible in complexity and realism by  integrating both synthetic and real-world data to address domain-shift issues, leverage physics-informed augmentations, ensure models are rooted in scientifically relevant tasks, and foster automation, explainability and interpretability to accelerate AI advancements in the field, and to develop a well-defined AI demonstrator for the wider AI community.

\subsection{(R7) Benchmarking and Standards}
\label{sec:Benchmarking}

Over the past two decades, machine learning methods have been tailored by the scientific community to extract optimal inference from complex datasets, optimize instruments, and improve reconstruction and pattern recognition processes. While the wide range of available techniques is beneficial, it can also be overwhelming for users facing new problems, leading to delays in developing optimal solutions.
The community survey response highlighted the need for common representative and reproducible benchmarks to provide a standardized way to measure progress and compare methods. 
For hosting datasets and models, platforms like Zenodo, OpenML, and Hugging Face offer centralized repositories, making it easier for researchers to access and share resources.
To address this, creating a knowledge base of tools suitable for common tasks in fundamental science will facilitate the adoption of new machine learning tools and reduce development time. Testing against common datasets also reduces development costs since benchmarking data products for validation are stored centrally and do not need to be regenerated by each development team. Therefore, community benchmarks will speed up proof-of-concept and prototyping activities and facilitate development of large scale models that tackle the challenges of live production data. 

Benchmarking and common datasets can also be used for community building, training and outreach. A potential approach is organizing community-wide challenges on carefully chosen, widely relevant, and complex problems. Examples include the 2014 Higgs and g2net gravitational wave challenges on the Kaggle platform, each  attracting submissions from over 1000 teams. The winning entries demonstrated the power of model pooling and cross-validation, showing the value of such competitions for the scientific community.

\textbf{Recommendation:}
Establish a dedicated effort to develop and maintain extensible benchmarks for various AI tasks in fundamental physics, such as event classification, parameter inference, tracking and anomaly detection. Support efforts to encourage researchers to share well-documented surrogate models to promote reusability and collaboration to drive innovation and standardisation in this area.

\subsection{(R8) Energy Efficiency}

The AI community now recognizes the environmental impact of large-scale model training (see e.g.~\cite{CarbonImpactArtificialIntelligence}), emphasizing the need to improve both algorithms and hardware use to reduce energy use and training time. 
It should also be noted that the use of AI inference can make data processing less resource-consuming as a whole, even when including training and hyperparameter search costs, by virtue of allowing to perform fast and efficient data reduction earlier in the data processing chain. 
Research in ``green AI,'' including insights from fundamental physics that aim to optimize performance metrics and hyperparameter scans for a reduced computing cost, is ongoing. 
Further strategies that can reduce emissions include choosing low-carbon-emission hardware (FPGAs, ASICs, and neuromorphic chips are designed to reduce the operational energy cost, mostly by reducing/optimizing memory access) and optimal compute locations can further reduce emissions.
Full lifecycle assessments for hardware should be considered as well, and our communities should connect to the computing sites where their training is ran, draw information from existing databases and experts, and encourage transparency in sharing information on where power and hardware resources come from.  
In addition, power-constrained environments, such as experiments at remote or extreme locations, can benefit from the development of novel neuromorphic readout schemes that leverage the natural time encoding of the processed information.

AI software/algorithmic sustainability must also be considered, alongside environmental sustainability.\footnote{
While we are not covering AI ethics and governance, we acknowledge that these topics have an impact with the design and application of AI algorithms in the fundamental sciences, and we refer to the reviews, overview and comment papers in Refs. \cite{LoPianoEthics,Thais:2024hzz,2024arXiv240714981R,2024arXiv240815244T}, as well as the AI Ethics and Governance in Practice Programme resources by the Turing Institute \cite{TuringEthicsGovernance}.} 
The definition of sustainable AI algorithms, as originally provided for research software \cite{DanKatzSustainability}, refers to algorithms that ``will continue to be available in the future, on new platforms, meeting new needs.''
Fast innovation in terms of AI theory and algorithm implementations in the fundamental sciences means that the lifecycle of any algorithm may be limited with respect to other types of software, reuse by others and reproducibility of the original algorithm's results is still a relevant issue for the field, and covered in the next section.   

\textbf{Recommendations:}
Investigate and adopt benchmarks that are suitable for fundamental sciences to raise awareness of the environmental impact of  large AI models. 
Consider collective mitigation strategies such as optimising widely used frameworks and models and their interfaces to existing software frameworks, as well as individual strategies that lead to minimal/acceptable performance loss. 
Cooperate with infrastructure and computing sites to minimise  carbon costs of compute-intensive AI tasks. 

\subsection{(R9) FAIR Principles}

The FAIR principles (Findable, Accessible, Interoperable, and Reusable) aim to improve data management and accessibility. These principles streamline data access, simplify the sharing of pre-trained models, and ensure resources are reusable across projects.
The framework of FAIR principles for both data \cite{wilkinson2016fair} and Research Software \cite{barker2022introducing} can be an appropriate starting point for improving collaboration, reproducibility, efficiency, and the visibility of data and research software.

Compliance to FAIR principles should be incorporated into publication practices, encouraging the publication of ML models, workflows, and associated example code. Despite challenges, a gradual implementation starting within experimental collaborations 
can lead to the development of a platform for ML models, workflows and code preservation similar to HEPData \cite{Maguire_2017}.
Policy measures should include providing resources for sustainable AI applications, with minimal bureaucratic overhead, and recognising FAIR compliance work in career progressions. Technical tools and standards must facilitate FAIR adherence, supported by national and international initiatives (see e.g.~\cite{Huerta_2023}, \href{https://everse.software}{EVERSE} and \href{https://fairos-hep.org}{FAIROS-HEP}). Building community awareness of FAIR principles from the undergraduate and graduate levels, and fostering alliances across scientific domains and regions, is vital for success.

\textbf{Recommendation:}
Develop activities aiming to integrate FAIR compliance into publication criteria and practices, recognise and incentivise the FAIR compliant work in policy and funding measures as well as career progression, build community awareness through training and collaboration, and support the development of technical tools and standards to facilitate the adoption of the FAIR principles.

\subsection{(R10) Training and Education}

Whilst there are many excellent examples of training and education programmes in specific settings, there is a lack of understanding on who should provide training, to what level of detail, and whether the training should sit within or across domains. Partly this is due to the field being fast-moving, with an absence of traditional learning (e.g.\ textbook-based) methods, which provide coverage from an introductory level to state-of-the-art algorithms and implementations, and a strong cross-disciplinary angle, with practices in computer science being remarkably different to those in fundamental science. Hence at present a significant amount of training is delivered via hands-on online courses, by interaction with the supervisor, or within a collaboration, all following slightly different standards.

Cross-disciplinary interactions are universally seen as being important. At the level of education and training, this can be implemented fairly easily by running cohort-based training modules in AI and data science, 
across disciplines
or via schools and workshops aimed at AI skills development rather than domain-specific questions. A European-wide effort in developing training courses---and keeping them up-to-date---would ensure a consistent quality and establish a common framework. Here one could take inspiration from the \href{https://datacarpentry.org/}{Data} and \href{https://software-carpentry.org/}{Software Carpentry courses}, for example.
A good example is the AI courses organized by the \href{https://erumdatahub.de/en/veranstaltungen/}{ErUM-Data Hub}, which demonstrate that funding for such training initiatives already exists.
By keeping this training not tied to a domain, ECRs (Early Career Researchers) will be enabled to interact naturally across disciplines, something which is rightly seen as very important for AI.    
Publishing codes and data with a paper is generally seen as important but is often only partly achieved due to the lack of clarity on what reproducibility entails. By embedding an agreed framework in the PhD training programme, it should become standard for ECRs.
One notable initiative is the \href{https://sites.google.com/view/lattice-virtual-academy}{{Lattice Virtual Academy} (LaVA)} \cite{Lava2024}, which was initiated at the European level via the \href{http://www.strong-2020.eu/}{STRONG-2020} Network. LaVA, hosted at ECT*, is a virtual platform for advanced e-learning in Lattice Field Theory, a computational method developed for nuclear and particle physics,  providing open-access, inclusive training to beginner and advanced students. In particular, it includes training on Reproducibility and Open Science \cite{athenodorou_2024_14222943}.

Interaction with industry, active in AI across the many sectors, may provide support in delivering the training from a slightly different perspective, and minimally lead to sponsorship of training events. A common model in AI is that industry provides placements for PhD students and postdocs. Here it is useful to point out that the topic of the placement is usually not linked to fundamental science, but rather to the application of AI and data science skills in a commercial environment, which will further develop the ECR skills set.

\textbf{Recommendation:}
Fund the development and organization of practical 
training courses and summer schools to equip researchers with the skills to implement open research and reproducibility requirements, incorporating examples and industry perspectives. Facilitate partnerships with industry to sponsor training events and provide placements for early-career and senior researchers, enhancing their AI and data science expertise while fostering connections between fundamental science and commercial applications.

\subsection{(R11) Interdisciplinary Collaboration}

The complexity and scale of AI research requires a collaborative approach that transcends disciplinary boundaries. Our survey highlighted the JENA community's strong interest in this approach, particularly for large-scale models. Effective AI solutions require domain-specific physics knowledge, necessitating close partnerships between physicists and AI specialists. This exchange ensures alignment between AI methods and scientific goals. This calls for structured partnerships where physicists contribute expertise on theoretical frameworks and experimental data, while AI specialists provide proficiency in model development and optimization techniques. 
While the primary focus should remain on physicists, AI specialists, software engineers, and HPC experts, collaborations could also extend to experts in fields such as social sciences, ecology, and public policy, where AI can address societal challenges. These broader partnerships foster cross-pollination of ideas and enhance the societal relevance of AI-driven physics research.
Furthermore, dedicated funding for interdisciplinary projects is crucial. 
Collaborative frameworks, like cross-institutional working groups, are needed to foster and promote knowledge exchange and better integrate AI into research workflows.
Establishing dedicated funding initiatives that explicitly target AI projects in scientific contexts could significantly enhance the capacity for innovation, allowing for the development of specialized models and computational tools that are fit-for-purpose in fundamental physics research. The integration of interdisciplinary collaboration with the existing efforts to enhance the education and training of researchers is also highly needed.
Establishing joint programs, workshops and courses, involving both AI specialists and domain scientists can help bridge knowledge gaps and build a common language between different disciplines. The objective is not only to develop AI tools but also to ensure these tools are accessible and usable by physicists and researchers in other scientific domains.

\textbf{Recommendation:}
Establish interdisciplinary research initiatives that bring together physicists, AI specialists, software engineers, HPC experts, and potentially experts from other related fields, to tackle large-scale projects. Provide dedicated funding to support cross-domain knowledge transfer through workshops, training programmes and open source collaboration. Invest in shared repositories and computing platforms to enable data sharing, modelling development and collaboration between different disciplines.

\section{(R12) Conclusion}
The rapid development of artificial intelligence is opening up new possibilities for fundamental physics and enabling breakthroughs in both research methods and scientific discovery. In this paper, comprehensive recommendations have been formulated to address the unique challenges in this field, with a focus on fostering collaboration, standardizing infrastructure and improving the scalability of AI applications.
At the heart of these efforts is a final recommendation that advocates the establishment of a dedicated organizational structure for the strategic coordination of AI investments in fundamental physics.
Such a structure, modeled on initiatives such as EuCAIF, would provide the necessary framework to focus resources, encourage interdisciplinary collaboration and accelerate the adoption of innovative AI technologies tailored to the specific needs of the field. 

By implementing these recommendations, fundamental physics can position itself at the forefront of AI innovation, set an example for interdisciplinary collaboration, and ensure sustainable progress in both scientific discovery and the broader societal impact of AI technologies. These efforts are particularly important for the success of future colliders and experiments that will require unprecedented levels of data processing, analysis and simulation. Integrating advanced AI techniques into these ambitious endeavors will not only enhance their scientific potential, but also provide a framework for tackling the  challenges of next-generation physics.

\textbf{Recommendation:}
Establish and support a dedicated organisational structure to coordinate strategic investments in AI for fundamental physics to accelerate the development and deployment of innovative AI technologies tailored to the specific challenges of the field. Existing initiatives like EuCAIF can serve as a model for such efforts.

\bibliographystyle{JHEP}

\begin{thebibliography}{10}

\bibitem{eucaif}
``{European Coalition for AI in Fundamental Physics (EUCAIF)}.'' \url{https://eucaif.org/}.

\bibitem{Albertsson:2018maf}
K.~Albertsson et~al., \emph{{Machine Learning in High Energy Physics Community White Paper}}, \href{https://doi.org/10.1088/1742-6596/1085/2/022008}{\emph{J. Phys. Conf. Ser.} {\bfseries 1085} (2018) 022008} [\href{https://arxiv.org/abs/1807.02876}{{\ttfamily 1807.02876}}].

\bibitem{Radovic2018}
A.~Radovic, M.~Williams, D.~Rousseau, M.~Kagan, D.~Bonacorsi, A.~Himmel et~al., \emph{Machine learning at the energy and intensity frontiers of particle physics}, \href{https://doi.org/10.1038/s41586-018-0361-2}{\emph{Nature} {\bfseries 560} (2018) 41}.

\bibitem{Boehnlein:2021eym}
A.~Boehnlein et~al., \emph{{Colloquium: Machine learning in nuclear physics}}, \href{https://doi.org/10.1103/RevModPhys.94.031003}{\emph{Rev. Mod. Phys.} {\bfseries 94} (2022) 031003} [\href{https://arxiv.org/abs/2112.02309}{{\ttfamily 2112.02309}}].

\bibitem{Cuoco:2020ogp}
E.~Cuoco et~al., \emph{{Enhancing Gravitational-Wave Science with Machine Learning}}, \href{https://doi.org/10.1088/2632-2153/abb93a}{\emph{Mach. Learn. Sci. Tech.} {\bfseries 2} (2021) 011002} [\href{https://arxiv.org/abs/2005.03745}{{\ttfamily 2005.03745}}].

\bibitem{Shanahan:2022ifi}
P.~Shanahan et~al., \emph{{Snowmass 2021 Computational Frontier CompF03 Topical Group Report: Machine Learning}},  2022 [\href{https://arxiv.org/abs/2209.07559}{{\ttfamily 2209.07559}}].

\bibitem{Boyda:2022nmh}
D.~Boyda et~al., \emph{{Applications of Machine Learning to Lattice Quantum Field Theory}},  in \emph{{Snowmass 2021}}, 2022 [\href{https://arxiv.org/abs/2202.05838}{{\ttfamily 2202.05838}}].

\bibitem{eurohpc_ai_factories}
{European Commission}, ``{AI Factories Funding Initiative}.'' \url{https://digital-strategy.ec.europa.eu/en/policies/ai-factories}, \url{https://ec.europa.eu/commission/presscorner/detail/en/ip_24_6302}.

\bibitem{CarbonImpactArtificialIntelligence}
P.~Dhar, \emph{The carbon impact of artificial intelligence.}, \href{https://doi.org/https://doi.org/10.1038/s42256-020-0219-9}{\emph{Nat. Mach. Intell} {\bfseries 2} (2020) 423}.

\bibitem{LoPianoEthics}
S.~Lo~Piano, \emph{{Ethical principles in machine learning and artificial intelligence: cases from the field and possible ways forward}}, \href{https://doi.org/https://doi.org/10.1057/s41599-020-0501-9}{\emph{Humanit Soc Sci Commun} {\bfseries 7} (2020) 9}.

\bibitem{Thais:2024hzz}
S.~Thais, \emph{{Physics and the empirical gap of trustworthy AI}}, \href{https://doi.org/10.1038/s42254-024-00772-7}{\emph{Nature Rev. Phys.} {\bfseries 6} (2024) 640}.

\bibitem{2024arXiv240714981R}
A.~{Reuel}, B.~{Bucknall}, S.~{Casper}, T.~{Fist}, L.~{Soder}, O.~{Aarne} et~al., \emph{{Open Problems in Technical AI Governance}},  \href{https://arxiv.org/abs/2407.14981}{{\ttfamily 2407.14981}}.

\bibitem{2024arXiv240815244T}
S.~{Thais}, \emph{{Misrepresented Technological Solutions in Imagined Futures: The Origins and Dangers of AI Hype in the Research Community}},  \href{https://arxiv.org/abs/2408.15244}{{\ttfamily 2408.15244}}.

\bibitem{TuringEthicsGovernance}
D.~Leslie, A.~Borda, A.~Perini and S.~Jayadeva, ``{AI Ethics and Governance in Practice Programme resources}.'' Available at: \url{https://aiethics.turing.ac.uk}, 2024.

\bibitem{DanKatzSustainability}
D.~Katz, ``{Defining Software Sustainability}.'' Available at: \url{https://danielskatzblog.wordpress.com/2016/09/13/defining-software-sustainability/}, 2016.
\newblock \href{https://doi.org/10.59350/naakc-7h373}{10.59350/naakc-7h373}.

\bibitem{wilkinson2016fair}
M.D.~Wilkinson, M.~Dumontier, I.J.~Aalbersberg, G.~Appleton, M.~Axton, A.~Baak et~al., \emph{The fair guiding principles for scientific data management and stewardship}, \href{https://doi.org/10.1038/sdata.2016.18}{\emph{Scientific Data} {\bfseries 3} (2016) 160018}.

\bibitem{barker2022introducing}
M.~Barker, N.P.~Chue~Hong, D.S.~Katz, A.-L.~Lamprecht, C.~Martinez-Ortiz, F.~Psomopoulos et~al., \emph{Introducing the fair principles for research software}, \href{https://doi.org/10.1038/s41597-022-01710-x}{\emph{Scientific Data} {\bfseries 9} (2022) 622}.

\bibitem{Maguire_2017}
E.~Maguire, L.~Heinrich and G.~Watt, \emph{{HEPData: a repository for high energy physics data}}, \href{https://doi.org/10.1088/1742-6596/898/10/102006}{\emph{Journal of Physics: Conference Series} {\bfseries 898} (2017) 102006}.

\bibitem{Huerta_2023}
E.A.~Huerta, B.~Blaiszik, L.C.~Brinson, K.E.~Bouchard, D.~Diaz, C.~Doglioni et~al., \emph{Fair for ai: An interdisciplinary and international community building perspective}, \href{https://doi.org/10.1038/s41597-023-02298-6}{\emph{Scientific Data} {\bfseries 10} (2023) }.

\bibitem{Lava2024}
M.-P.~Lombardo et~al., \emph{{LaVA -- Lattice Virtual Academy}},  2024.
\newblock \href{https://www.ectstar.eu/virtual-platforms/}{www.ectstar.eu/virtual-platforms/}.

\bibitem{athenodorou_2024_14222943}
A.~Athenodorou, E.~Bennett and J.~Lenz, \emph{{Lattice Virtual Academy: Reproducibility and Open Science}},  2024.
\newblock \href{https://doi.org/10.5281/zenodo.14222943}{10.5281/zenodo.14222943} and {\href{https://sites.google.com/view/lattice-virtual-academy/topics/open-science?authuser=0}{LaVA module}}.

\end{thebibliography}

\providecommand{\href}[2]{#2}\begingroup\raggedright\endgroup

\end{document}